# Fully Convolutional Generative Machine Learning Method for Accelerating Non-Equilibrium Green's Function Simulations

Preslav Aleksandrov
Device Modelling Group
James Watt School of Engineering
University of Glasgow, UK
preslav.aleksandrov@glasgow.ac.uk

Ali Rezaei
Device Modelling Group
James Watt School of Engineering
University of Glasgow, UK
ali.reazaei@glasgow.ac.uk

Nikolas Xeni
Device Modelling Group
James Watt School of Engineering
University of Glasgow, UK
nikolas.xeni@glasgow.ac.uk

Tapas Dutta
Device Modelling Group
James Watt School of Engineering
University of Glasgow, UK
tapas.dutta@glasgow.ac.uk

Asen Asenov
Device Modelling Group
James Watt School of Engineering
University of Glasgow, UK
asen.asenov@glasgow.ac.uk

Vihar Georgiev
Device Modelling Group
James Watt School of Engineering
University of Glasgow, UK
vihar.georgiev@glasgow.ac.uk

*Abstract*— **This work describes a novel simulation approach that combines machine learning and device modeling simulations. The device simulations are based on the quantum mechanical non-equilibrium Green's function (NEGF) approach and the machine learning method is an extension to a convolutional generative network. We have named our new simulation approach ML-NEGF and we have implemented it in our in-house simulator called NESS (nano-electronics simulations software). The reported results demonstrate the improved convergence speed of the ML-NEGF method in comparison to the 'standard' NEGF approach. The trained ML model effectively learns the underlying physics of nano-sheet transistor behaviour, resulting in faster convergence of the coupled Poisson-NEGF simulations. Quantitatively, our ML-NEGF approach achieves an average convergence acceleration of 60%, substantially reducing the computational time while maintaining the same accuracy.**

*Keywords— machine learning, neural network, autoencoder, device simulations and modeling, non-equilibrium Green's function (NEGF), TCAD device modeling, nanowires.*

## I. INTRODUCTION

The silicon nanowire and nanosheet transistors have a wide spectrum of promising applications [1], such as current field-effect transistors [2] and photovoltaics [3]. Moreover, the state-of-the-art CMOS technologies are based on single or stacked configurations of nanosheet or nanowire architectures [4]. However, despite the recent advances in technology, there is still room to improve the fabrication process and to optimise device performance, for example, by reducing power consumption and reducing device-to-device variability during the fabrication process.

From a practical point of view, simulations and modelling transistors are the most time-efficient and cost-effective approach to evaluate the performance and the output characteristics of transistors. The aim is to have a simulation platform that is fast, accurate, and reliable in order to aid the improvement of device design, predict device performance (current-voltage characteristics) and extract important Figures of Merit (FoM).

The main aim of this work is to investigate the possibility of significantly improving or even replacing numerical

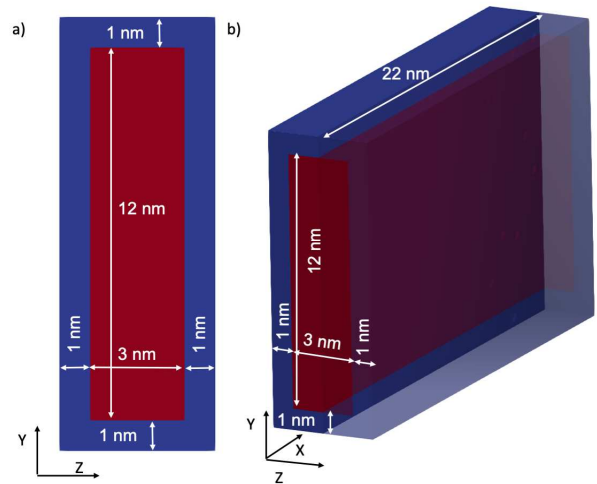

Fig. 1. Diagram of the n-type silicon (Si) nanosheet transistor with channel cross section of 3 nm x 12 nm (a) and length of full device 22 nm (b). Gate length is 16 nm and the source/drain have length of 3 nm each. The channel doping is 1e16 $cm^{-3}$ and the contacts (source and drain have) are 1e20 $cm^{-3}$. The oxide is $SiO_2$ with thickness of 1nm everywhere around the device.

Technology Computer-Aided Design (TCAD) device simulations with a convolutional autoencoder (CAE) [5] [6] [7]. To test our idea, we have developed a new simulation approach based on the combination of TCAD and machine learning methods. The current state of the art of TCAD simulations is based on the Non-equilibrium Green's Function (NEGF) formalism that can capture the quantum mechanical physical effects, such as confinement and carrier tunnelling in ultra-scaled transistor (with channel lengths that are shorter than 10 nm). To enhance the capabilities of the NEGF method and decrease the computational time of our in-house Nano Electronics Simulation Software (NESS) [8], we combine machine learning with our existing NEGF simulator implemented in NESS. We have called the new simulation approach ML-NEGF.

Our results show the potential of using the ML-NEGF methodology to significantly reduce the device simulation computational cost without compromising the accuracy of physical results deriving from the 'standard TCAD' simulations.

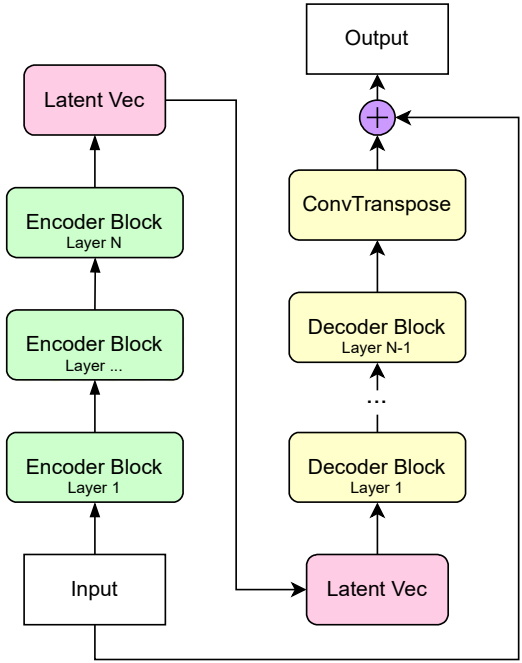

*Fig. 2. A diagram of the convolutional autoencoder structure. It consists of N encoder blocks and N-1 decoder blocks and a final convolutional layer.*

## II. DEVICE STRUCTURE

To test our new simulation ML-NEGF methodology we have designed a device transistor structure that corresponds to the most advanced technologies of 3 nm node and beyond. Fig. 1 shows the nanosheet transistor geometry created using the NESS structure generator. The gate is all around the channel and the channel length is $L_{Ch}$ = 16 nm, with source and drain lengths of 3 nm each, hence the total length of the device is 22 nm. The channel cross-section is rectangular with dimensions of 3 nm x 12 nm, the oxide material is $SiO_2$ and its thickness is 1 nm. The channel doping is 1e15 $cm^{-3}$ and the source and drain regions have doping of 1e20 $cm^{-3}$.

## III. SIMULATIONS METHODOLOGY

All numerical simulations in the work are performed by utilising the NEGF simulator implemented in our in-house code NESS [8]. The NEGF implementation is based on the effective mass approximation. Our NEGF solver can compute ballistic and scattering transport in various devices and materials. In this paper, we have used the ballistic version of the NEGF solver to test our idea. However, we would like to emphasise that our methodology is valid even if we use the simulations that include the electron-phonon and surface roughness scattering mechanisms in the active region of the device.

The NEGF solver is linked to a 3D Poisson solver and both solvers are connected in a self-consistent loop. The effective-mass Hamiltonian, and correspondingly the NEGF solver, requires potential as an input that is provided by the Poisson solver. Correspondingly the Poisson solver requires a charge that is provided by the NEGF solver. Also, the NEGF formalism allows to compute device characteristics, such as current-voltage curves ($I_D$-$V_G$ and $I_D$-$V_D$). From the $I_D$-$V_G$ curve, we can extract important FoM, such as OFF-current ($I_{OFF}$) and ON-current ($I_{ON}$), subthreshold slope (SS) and voltage threshold ($V_{TH}$). In previous papers, we have shown that it is possible to train a neural network (NN) by using as input date key figures of merit such as subthreshold slope, drive current, leakages current to predict another key parameter such as voltage threshold [9].

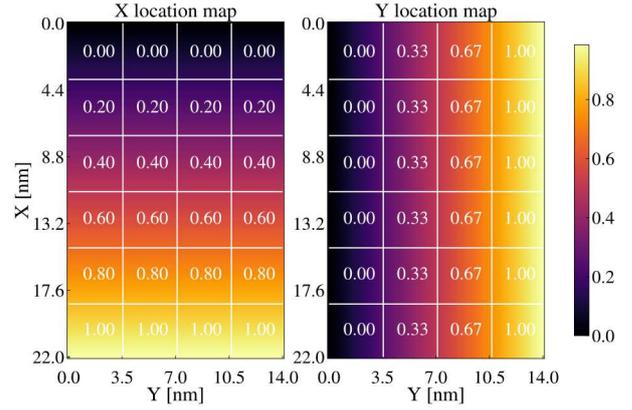

*Fig. 3. Location maps in X and Y directions. X is the transport direction with length of 22 nm and Y is the longer cross-section with is 14 nm long. The colour map shows the location of the kernel in relation to the model. Numbers and lines show the rough splitting of data.*

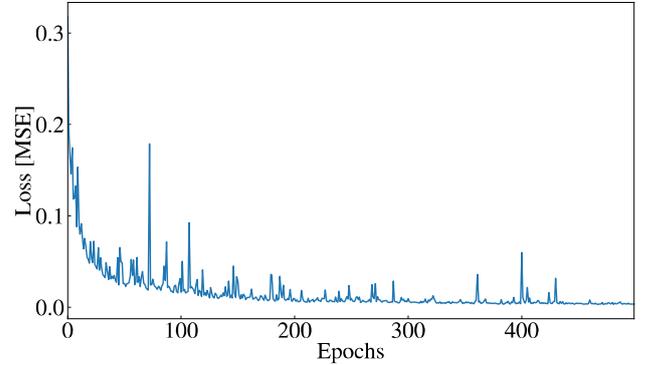

*Fig. 4. Validation Loss of the model and a function of the Epochs (steps) of the convolutional autoencoder.*

In this paper, we have utilised a machine-learning model inspired by denoising autoencoders. The ML model is shown in Fig. 2. The model's architecture is based on a convolutional denoising autoencoder network augmented by methods stemming from transformer networks. The basic structure was chosen to be fully convolutional as this guarantees model generality and improves robustness to different device geometries. The augmentations, borrowed from transformer networks, are the inclusion of location encodings in the initial input. A residual connection between the input and output of the model was also introduced to reduce the solution domain to the change between initial and final NEGF-Poisson iterations. The output of the models consists of a single channel matrix, which represents a normalised field of potential or charge. The output is assumed to be normalised with respect to the mean and deviation of the input.

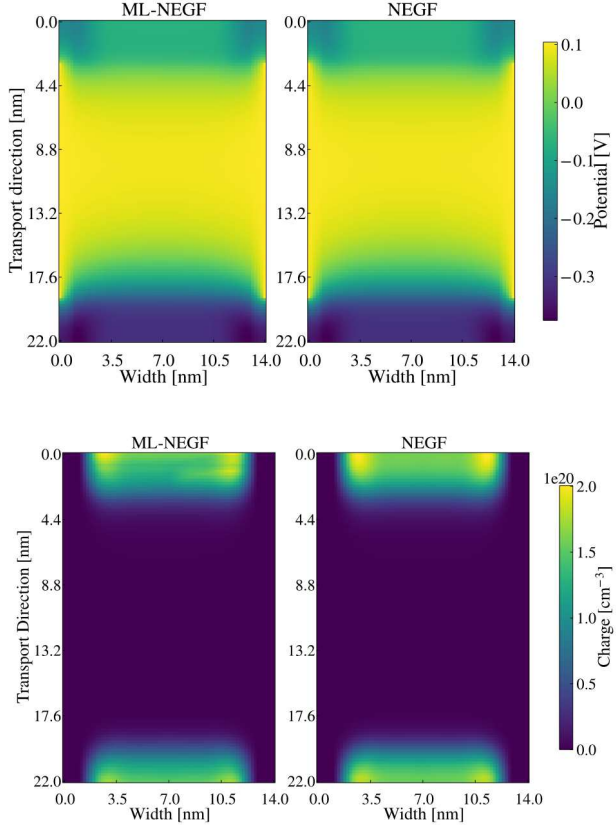

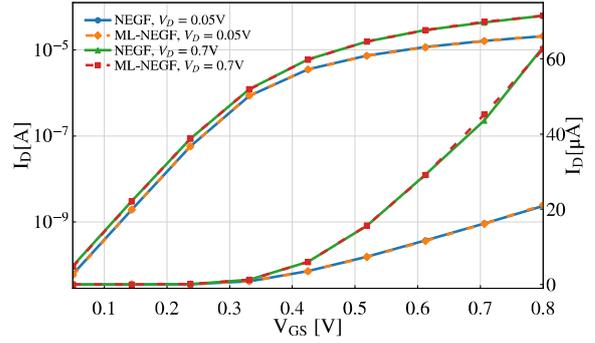

*Fig 6. Comparison of the current voltage characteristic ($I_D$-$V_G$ curves) for both the ML-NEGF and NEGF methods, as a function of the gate bias ($V_{GS}$) at fix drain bias (Low =0.05V and High = 0.7V).*

*Fig. 5. Comparison of charge distribution (top row) and potential distribution (bottom row) in a XY plane along the transport direction, between the ML-NEGF and 'standard' NEGF simulations. The charge has the highest value in the source and drain region and the highest potential in the middle of the channel.*

The model can be constructed using *N* number of encoder and decoder blocks, to extract a latent space representation of the input and apply relevant transformations in the decoder section. Each block consists of a convolution, batch normalisation, dropout and activation function. We chose the LeakyRelu function as the activation function in the encoder section as it has a high gradient. One could also introduce residual connection within the network to counteract the diminishing gradient problem. However, due to the small depth of the network, *N*=3, this was not implemented.

A set of location matrices, presented in Fig. 3, was added to hint the location of the kernel to the model. The location encodings were artificially generated, by making a gradient map between 0 and 1 in each of the basis directions (X, Y, Z). The range between 0 and 1 was chosen to maintain generality.

The main aim is to train an ML model to predict the difference between the 3D spatial charge and the potential distributions of the first and final Poisson-NEGF self-consistency iterations inside the whole of the device. As an input to the autoencoder-accelerated ML-NEGF method, we provided the charge and potential obtained from the initial ballistic Poisson-NEGF iteration of the self-consistent loop. The input of the model is a 7-channel image generated from information produced by an initial NEGF iteration. The first two channels are a normalised 2D image slice of potential and of charge in logarithm scale. The next two channels are the drain and gate voltages. The final 3 channels, two of which are shown in Fig. 3, are the location maps used by the kernel. Each square represents the chosen value for a specific device location and the colour bar gives the heat map of those values.

The model output can then be used as an input to the 'standard' NEGF simulations to reduce the number of self-consistent Poisson-NEGF iterations, which leads to a significant reduction of the simulation time. The fully trained model can be examined as a kernel-based analytical representation of the NEGF solver. The solution of which is the forward pass of the ML model. Computationally, the cost of this is negligible if compared to the cost of utilising the NEGF solver. Therefore, this method, once trained, is computationally efficient and can be used to accelerate NEGF simulations.

## IV. SUMULATION RESULTS

In order to validate our ML-NEGF model, input and target data was generated by our 'standard' NEGF simulation. The obtained data was divided into two sets: training and testing. The training set is 70%, and the testing set is 30% of the full data. The training set is used to train the ML model. The loss characteristic obtained from the training process is shown in Fig. 4.

Fig. 4 shows the evolution of the mean square error (MSE) as a function of the epochs (training steps). The model was trained for 500 epochs, where it reached a point of saturation. The number of 500 epochs was discovered empirically. The training loss follows a typical trend, where it shows significant oscillation and an exponential decrease in around 100 epochs, followed by a reduction in training speed between 100 and 400 epochs. After 400 epochs the MSE value has saturated (0.02 MSE).

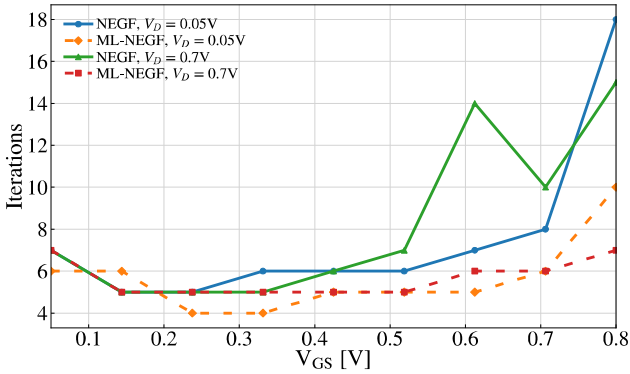

*Fig. 7. Comparison of the simulations of self-consistent iterations between ML-NEGF and NEGF (ballistic), as a function of the gate bias ($V_{GS}$).*

Fig. 5 shows the comparison between the 2D charge and potential distribution in the middle of the channel, along the transport direction, for the ML-NEGF simulations and the target scalar fields produced by the 'standard' NEGF method. Consistent with the device structure and the doping profile along the device, the charge value is the highest in the source and the drain region and the smallest in the channel section of the device. Also, consistent with the device physics, at low gate biases voltages, below 0.4V, the potential has the highest value in the middle of the channel. From the results in Fig. 5, it can be concluded that the charge and the potential distribution obtained from the ML-NEGF method are identical to those extracted from NEGF. Hence, it can be concluded that our convolutional NN is indeed well trained.

In Fig. 6, we have plotted and compared the $I_D$-$V_G$ curves for both the ML-NEGF and 'standard' NEGF approach, at low ($V_D$=0.05V) and high ($V_D$=0.7V) drain bias. From the results in Fig. 6, it is evident that both methods produced identical $I_D$-$V_G$ curves and, hence, the FoMs, that can be extracted, will be also identical. The results in Fig. 5 and Fig. 6 show that our NN used in the ML-NEGF method can reproduce not only physical prosperities but also key device characteristics.

Once the ML model is trained, we wanted to evaluate and compare the convergence behaviour for both cases. Fig. 7 shows the number of self-consistent interactions as a function of gate voltage ($V_{GS}$), at low ($V_D$=0.05V) and high ($V_D$=0.7V) drain bias. From the data in Fig. 7, it can be concluded that overall, the ML-NEGF method requires a smaller number of iterations in comparison to the NEGF method. Specifically, at up to 0.2 $V_{GS}$ both methods have almost identical iterations. However, when the $V_{GS}$ has values above 0.2V, the ML-NGEF simulations (see the red and orange curves in Fig. 7) show a consistently lower iterations number than the 'standard' NEGF method. For example, the difference between both methods is well pronounced at $V_{GS}$=0.8V. At low drain bias ($V_D$=0.05V), ML-NEGF (orange curve) converges in 10 iterations, while the conventional NEGF method (blue curve) needs 18 iterations. At high ($V_D$=0.05V) drain bias, ML-NEGF (red curve) requires 7 steps and the NGEF method (green curve) converges after 15 steps. Hence, in both cases the ML-NEGF approach achieves an average convergence acceleration of 60%, substantially reducing the computational time, while maintaining the same accuracy.

## V. Conclusions

In this work, we have reported a combined machine learning and device simulation computational approach that allows us to simulate the device characteristics (current-voltage) of Si nanosheet transistors. Our machine learning method is based on a convolutional neural network and autoencoder architecture. Results obtained from the ML-NEGF approach led to the following conclusions.

Firstly, using the autoencoder-accelerated ML-NEGF method instead of the standard TCAD (NEGF) simulations in principle could significantly decrease the computational time and shorten the research and development process. For example, the ML-NEGF approach achieves an average convergence acceleration of 60%, while maintaining the same accuracy. Secondly, our autoencoder-accelerated ML-NEGF method can reproduce not only the device characteristics but also 3D charge density and potential distribution in the whole device. Lastly, similar ML based approach can be used to describe material properties, such as resistance in metal nanowires that cannot be described by non-parametric methods, such as a general linear model. However, it needs to be noted that the predictivity of the ML-NEGF method can be improved even further by providing more data, using different pre-processing schemes and attempting alternative network architectures. Indeed, all these options are currently under investigation.

ACKNOWLEDGMENT

This research was funded by the Engineering and Physical Sciences Research Council (EPSRC), through Grant No. EP/S001131/1 and EP/P009972/1. This project has also received funding from the EPSRC Impact Acceleration Account scheme under Grant Agreement No. EP/R511705/1 (Nano-Electronic Simulation Software (NESS)—creating the first open source TCAD platform in the world and Fast Track - Development boost for the Device Modelling group opensource NESS computational framework).